\documentclass[aps,preprint,amsmath,amssymb,superscriptaddress,nofootinbib]{revtex4}

\usepackage{graphicx}
\begin{document}

\title{Some surprises in the neutrino cross sections associated with neutrino spin}

\author{\.{I}nan\c{c}  \c{S}ahin}
\email[]{inancsahin@ankara.edu.tr}
 \affiliation{Department of
Physics, Faculty of Sciences, Ankara University, 06100 Tandogan,
Ankara, Turkey}

\begin{abstract}

It is generally assumed that neutrino masses can be neglected to a
high degree of approximation in cross section calculations. This
assumption seems very reasonable since the neutrino masses are
extremely small and the neutrinos are ultrarelativistic fermions at
the energy scales of current experiments. Consequently, in cross
section calculations in the Quantum Field Theory, the Standard Model
neutrinos are frequently assumed to be described by $100\%$ negative
helicity states. This assumption is true in a sense that in the
Standard Model processes the positive helicity states can be safely
neglected for ultrarelativistic neutrinos. On the other hand, the
assumption tacitly assert that the neutrino fields are completely
longitudinally polarized, i.e., the contribution to the cross
section coming from transverse polarization can be neglected. We
show that this tacit assertion is not correct. Although the Standard
Model cross section for a neutrino with positive helicity goes to
zero as $m_\nu\to 0$, the cross section for a neutrino with
transverse polarization remains finite in that limit. Thus the
contribution coming from transverse polarization cannot be neglected
even in the ultrarelativistic/zero-mass limit. We examine the
consequences of this fact and deduce that it has some unexpected
results in the neutrino cross sections.

\end{abstract}


\maketitle

\section{Introduction}

According to the Standard Model (SM) of particle physics the
neutrinos couple minimally to other SM particles only through $V-A$
type vertex and hence the interaction project out the "left" chiral
component of the neutrino field. Consequently, all interacting
neutrinos in the SM can be accepted to be left chiral which can be
written mathematically as $\frac{1}{2}(1-\gamma_5)u_\nu(p)=u_\nu(p)$
where $u_\nu(p)$ is the spinor for the neutrino \footnote{In this
paper only Dirac neutrinos and their SM interactions have been
considered.}. It is well known that massless fermions are completely
longitudinally polarized \cite{Wigner}. They are described by pure
helicity states which coincide with chirality eigenstates. If the
mass of the fermion is zero then its positive and negative helicity
states coincide with right-handed and left-handed chirality
eigenstates respectively. Since all SM neutrinos are left chiral,
massless neutrinos must be described by $100\%$ negative helicity
states. On the other hand, as we know from experimental results
obtained in Super-Kamiokande and Sudbury Neutrino Observatory that
neutrinos oscillate and they cannot be massless
\cite{Fukuda:1998mi,Ahmad:2001an}. Although neutrinos are not
massless they possess very tiny masses and hence they are
ultrarelativistic at the energy scales of current experiments.
Consequently, during cross section calculations it is generally
assumed (except for some direct neutrino mass measurement
experiments) that neutrino masses can be neglected and the neutrinos
are described by $100\%$ negative helicity states. Ignoring neutrino
masses is an approximation which is believed to be valid with a high
degree of accuracy for energies much greater than the neutrino mass.
On the contrary, we will show in this paper that the approximation
is not as accurate as expected even in the zero-mass limit.

The crucial point which is generally skipped in the literature is
that the solutions of the free Dirac equation describing a general
spin orientation have a discontinuity at the point $m=0$. If we take
the zero-mass limit of the spinor $u^{(s)}(p)$ describing a general
spin orientation we do not get, in general, its value evaluated at
$m=0$, i.e., $\lim_{m\to0}u^{(s)}(p)\neq u^{(s)}(p)\mid_{m=0}$
\cite{Sahin:2016bjs}. According to the seminal work of Wigner
\cite{Wigner}, strictly massless fermions are longitudinally
polarized and described solely by helicity eigenstates. However if
the fermion has a non-zero mass (no matter how small it is), then it
is allowed to have an arbitrary spin orientation which is different
from longitudinal direction. It is quite surprising that the
transverse polarization does not disappear in the zero-mass limit
but it vanishes instantly at the point $m=0$ \cite{Sahin:2016bjs}.
This behavior is the origin of the discontinuity of the free Dirac
solutions with general spin. If we restrict ourself to special type
of Dirac solutions, namely helicity states we observe that the
helicity states converge to the chirality eigenstates in the
zero-mass limit and we do not encounter any discontinuity at $m=0$.
However, the zero-mass behavior observed from helicity states is not
valid in general. Although the helicity states converge to the
chirality eigenstates in the zero-mass limit, a spinor with
arbitrary spin orientation does not necessarily result in a
chirality eigenstate in that limit. For instance, the spinor with
transverse polarization (relative to the direction of momentum) is
always given by a mixed chirality eigenstate and hence does not
converge to one of the chirality eigenstates left-handed or
right-handed even in the zero-mass limit \footnote{The explicit
expressions for Dirac spinors describing a general spin orientation
and their behavior in the zero-mass limit can be found in
Ref.\cite{Sahin:2016bjs} in detail.}. The discontinuity of the free
Dirac spinors at $m=0$ induces a similar discontinuity in the cross
sections. If we calculate the cross section for a neutrino with mass
$m_\nu$ and then take its $m_\nu\to0$ limit what we get is different
from the cross section in which the neutrino is initially assumed to
be massless, i.e., $\lim_{m_\nu\to0} \sigma(m_\nu)\neq \sigma(0)$.
As a result of this discontinuous behavior, neglecting neutrino
masses in the cross section is not a good approximation even though
neutrino masses are extremely small compared to the energy scale of
the processes that we consider.

The organization of the paper is as follows. In section II we review
the free Dirac spinors describing a general spin orientation and
their discontinuous behavior at $m=0$. In section III-A we present
cross section calculations in some generic SM processes, assuming
that neutrinos are massless. In section III-B the cross section
calculations are performed for massive neutrinos but the mixing
between different mass eigenstates is omitted for simplicity. In
section III-C, a more realistic situation is considered where both
neutrino masses and mixing are taken into account. In the
conclusions section (section IV) we summarize the results that we
obtain and discuss briefly some of its implications.

\section{Zero-mass discontinuity of the Dirac spinors}

Let us review shortly the free Dirac spinors describing a general
spin orientation. Assume that in the rest frame of the fermion, its
spin is quantized along the direction defined by the unit vector
$\vec n$. Then, in the rest frame we can write the following
eigenvalue equations
\begin{eqnarray}
\label{non-relativisticspin} (\vec n \cdot\vec S)
u^{(\uparrow)}_{RF}=+\frac{1}{2} u^{(\uparrow)}_{RF},\;\;\;\;(\vec n
\cdot\vec S)
u^{(\downarrow)}_{RF}=-\frac{1}{2}u^{(\downarrow)}_{RF};\;\;\;\;
\end{eqnarray}
where $\vec S=\frac{1}{2}\left(
                              \begin{array}{cc}
                                \vec\sigma & 0 \\
                                0 & \vec\sigma \\
                              \end{array}
                            \right)$
is the non-relativistic $4\times4$ spin matrix. The eigenvectors
(rest spinors) which correspond to the eiegenvalues $+1/2$ and
$-1/2$ are called spin-up ($\uparrow$) and spin-down ($\downarrow$)
spinors respectively. The spinor for a moving fermion can be
obtained by applying a Lorentz boost to the spinor at rest.  Suppose
that $S'$ frame is moving along negative $z$-axis with relative
speed $v$ with respect to the rest frame of the fermion $S$. Then
the observer in the $S'$ frame sees a moving fermion with
four-momentum $p^\mu=(E,\vec p)=(E,0,0,p_z)$ and four-spin
\cite{GreinerRQM,GreinerQED}
\begin{eqnarray}
\label{fourspin} s^\mu=L^\mu_\nu
\left(s^\nu\right)_{RF}=\left(\frac{\vec{p}\cdot\vec{n}}{m},\vec{n}+\frac{\vec{p}\cdot\vec{n}}{m(E+m)}\vec{p}\right)
\end{eqnarray}
where $L^\mu_\nu$ is the Lorentz transformation tensor and
$\left(s^\nu\right)_{RF}=(0,\vec n)$ is the spin vector defined in
the rest frame of the fermion. Without loss of generality, choose
$\vec{n}=\sin\theta\; \hat x+ \cos\theta\; \hat z$, i.e., $\vec{n}$
is in the $z$-$x$ plane which makes an angle $\theta$ (polar angle)
with respect to the $z$-axis. Then according to an observer in $S'$,
the spin-up ($\uparrow$) and spin-down ($\downarrow$) spinors
describing a general spin orientation are given by
\cite{Sahin:2016bjs}
\begin{eqnarray}
\label{spinup}
u^{(\uparrow)}(p)=\cos{\left(\frac{\theta}{2}\right)}\;u^{(+)}(p)+\sin{\left(\frac{\theta}{2}\right)}\;u^{(-)}(p)\\
\label{spindown}
u^{(\downarrow)}(p)=\cos{\left(\frac{\theta}{2}\right)}\;u^{(-)}(p)-\sin{\left(\frac{\theta}{2}\right)}\;u^{(+)}(p)
\end{eqnarray}
where $u^{(+)}(p)$ and $u^{(-)}(p)$ represent positive and negative
helicity spinors which correspond to the special spin orientation
(special orientation of the spin quantization axis) $\vec
n=\frac{\vec p} {|\vec p|}$ or equivalently $\theta=0$. It is
obvious from equations (\ref{non-relativisticspin}) or
(\ref{spinup}) and (\ref{spindown}) that the spin-up and spin-down
spinors are interchanged under the transformation $\vec n\to -\vec
n$ or equivalently $\theta \to \pi+\theta$. Finally, let us stress
the simple but crucial point which is at the heart of the analysis
presented in this paper. \emph{The angle $\theta$ that appears in
equations (\ref{spinup}) and (\ref{spindown}) is not a dynamical
variable. It does not depend on the relative velocity between the
frames $S$ and $S'$. It is the angle measured in the frame in which
the particle is at rest. Hence, $\theta$ is not affected by
relativistic aberration.} The angle $\theta$ resembles the term
"proper time" which is a frame-independent quantity. Due to this
resemblance, we will call it "proper angle". We should also remind
the fact that when we talk about the spin orientation of a moving
fermion, we mean the orientation of the spin quantization axis $\vec
n$ in the \emph{rest frame} of the particle. Therefore the spinors
(\ref{spinup}) and (\ref{spindown}) for a general spin orientation,
describe a fermion which in its rest frame its spin is quantized
along $\vec{n}=\sin\theta\; \hat x+ \cos\theta\; \hat z$.

Now we are ready to discuss the discontinuous behavior of the Dirac
spinors at $m=0$. Since the expressions (\ref{spinup}) and
(\ref{spindown}) for spinors with general spin orientation are
obtained by means of a Lorentz transformation from the rest frame of
the fermion $S$ to a moving frame $S'$, they remain valid for every
value of the relative speed $v$ that satisfies $|v-c|<\epsilon$
where $\epsilon$ is infinitesimal. Consequently, the zero-mass
($m\to0$) or ultrarelativistic ($v\to c$) limit of the spinors
$u^{(\uparrow)}(p)$ and $u^{(\downarrow)}(p)$ exists and given by
\begin{eqnarray}
\label{limitspinup} \lim_{m\to 0}u^{(\uparrow)}(p)=\cos{\left(\frac{\theta}{2}\right)}\;u^{(R)}(p)+\sin{\left(\frac{\theta}{2}\right)}\;u^{(L)}(p)\\
\label{limitspindown} \lim_{m\to
0}u^{(\downarrow)}(p)=\cos{\left(\frac{\theta}{2}\right)}\;u^{(L)}(p)-\sin{\left(\frac{\theta}{2}\right)}\;u^{(R)}(p)
\end{eqnarray}
where $u^{(R)}(p)=\lim_{m\to 0}u^{(+)}(p)=u^{(+)}(p)\mid_{m=0}$ and
$u^{(L)}(p)=\lim_{m\to 0}u^{(-)}(p)=u^{(-)}(p)\mid_{m=0}$ are the
right-handed and left-handed chirality eigenstates. On the other
hand, the Lorentz group is non-compact and the parameter space of
the Lorentz group does not contain the point $v=c$. Therefore, we
cannot perform a Lorentz transformation to the rest frame of a
massless particle. In other words, massless particles do not have a
rest frame. Consequently, the expressions (\ref{limitspinup}) and
(\ref{limitspindown}) become invalid for strictly massless
particles. In the case of massless fermions, we should employ the
little group analysis of Wigner \cite{Wigner}. According to Wigner
massless particles are described by $E(2)$-like little group and
that their spin orientations other than parallel or antiparallel to
the direction of momentum are not allowed. Hence, massless fermions
must be completely longitudinally polarized and described by pure
helicity states which coincide with chirality eigenstates. We
observe from equations (\ref{limitspinup}) and (\ref{limitspindown})
that the zero-mass limit of the spinors for a general spin
orientation are not equal to a chirality eigenstate unless
$\theta=0$ or $\pi$. Therefore the spinors $u^{(\uparrow)}(p)$ and
$u^{(\downarrow)}(p)$ have a discontinuity at $m=0$ that is
$\lim_{m\to0}u^{(s)}(p)\neq u^{(s)}(p)\mid_{m=0}$ for
$\theta\neq0,\pi$ where $u^{(s)}(p)\mid_{m=0}$ is either
$u^{(R)}(p)$ or $u^{(L)}(p)$. In the case of SM neutrinos
$u^{(s)}(p)\mid_{m=0}=u^{(L)}(p)$.

\section{Neutrino cross section for general spin orientation}
\subsection{Massless case}

 In the SM of particle physics, neutrinos
interact through the weak interaction. Hence any SM process which
contains the neutrinos involve $W$ or/and $Z$ boson exchange. The
former generates charged current and the latter generates neutral
current neutrino interactions. In both of the cases, the interaction
is proportional to the left chirality projection operator $\hat
L=\frac{1}{2}(1-\gamma_5)$. Hence, the neutrinos must be left-handed
chiral in interactions. This fact is always true in the SM,
independent to whether the neutrinos are massless or not. Assume
that neutrinos are strictly massless. In this case, the neutrinos
must also be described by a pure negative helicity state. This is
evident since massless fermions are completely longitudinally
polarized, and that their positive and negative helicity eigenstates
coincide with right-handed and left-handed chirality eigenstates.
Possibly because of this reason, sometimes the terms "left-handed"
and "negative helicity" are used interchangeably in the literature
for massless neutrinos, although there are some differences in their
meaning. However one should be very careful in the case of massive
neutrinos and does not use these terms instead of each other even
though neutrino masses are extremely small.\footnote{Sometimes the
terms "left-handed" and "right-handed" are used for the eigenstates
of the helicity instead of chirality. This is a matter of convention
but the important thing is not to confuse the eigenstates of the
helicity and chirality. In this paper we use the terms "left-handed"
and "right-handed" for the eigenstates of the chirality and
"negative helicity" and "positive helicity" for the eigenstates of
the helicity.}

Let us first assume that neutrinos are strictly massless and their
flavor and mass eigenstates coincide. Consider single neutrino
production and absorption processes $ab\to \nu c$ and $\nu a'\to
b'c'$ where $a,b,c,a',b',c'$ are charged fermions. The tree-level
amplitudes for these processes can be written in the form
\begin{eqnarray}
\label{amplitudechargedcurrent1}
M=g\left({J_C}^\alpha\;{J'_C}_\alpha\right)\mid_{m_\nu=0}
\end{eqnarray}
where ${J_C}^\alpha$ is the charged current that contains the
neutrino field and ${J'_C}^\alpha$ is the charged current for
charged fermions and $g$ is some constant. In writing equation
(\ref{amplitudechargedcurrent1}) we assume that the $W$ propagator
can be approximated as $\frac{(g_{\mu\nu}-q_\mu
q_\nu/m_w^2)}{q^2-m_w^2}\approx-\frac{g_{\mu\nu}}{m_w^2}$. The
explicit form of the charged neutrino current is given by
\begin{eqnarray}
\label{chargedcurrent-zeromass1} {J_C}^\alpha&&=\left[\bar u_\nu
\gamma^\alpha
\hat L u_\ell\right]\;\;\;\;\text{for the process}\;\;ab\to \nu c\\
\label{chargedcurrent-zeromass2} {J_C}^\alpha&&=\left[\bar
u_{\ell\;'} \gamma^\alpha \hat L u_\nu\right]\;\;\;\;\text{for the
process}\;\;\nu a'\to b'c'
\end{eqnarray}
where $\hat L=\frac{1}{2}(1-\gamma_5)$ is the left chirality
projection operator. The unpolarized cross section is proportional
to the squared amplitude which is averaged over initial and summed
over final spins. In the case of single neutrino production, the sum
over final neutrino spin gives just one term with $s_\nu=-1$ that
corresponds to negative helicity. Hence, for $m_\nu=0$ the produced
neutrinos are \emph{completely longitudinally polarized} \footnote{
The statement "completely longitudinally polarized" is used in the
meaning that the only possible spin orientation is the one which is
parallel or anti-parallel to the direction of momentum.} and
described by a state with $100\%$ negative helicity. In the case of
single neutrino absorption, \emph{we do not perform an average over
initial neutrino spins and omit the factor of 1/2 coming from spin
average of initial neutrinos}. Omitting initial neutrino spin
average is based on the assumption that all neutrinos in the SM are
described by completely longitudinally polarized negative helicity
states. This assumption is obviously true for massless neutrinos.
The neutrinos which enter neutrino absorption processes should be
produced through some production processes. If the neutrinos are
strictly massless, then all produced neutrinos through a SM process
are indeed completely longitudinally polarized and described by a
state with $100\%$ negative helicity. Nevertheless, as we will see
in the next subsection surprisingly, the assumption underestimates
the cross section for neutrinos with non-zero mass even though
neutrino masses are very tiny.

Now let us consider another simple process the neutrino scattering
from a charged fermion, $\nu a\to \nu a$. Depending on the type of
the charged fermion $a$, the process may contain only neutral or
both neutral and charged neutrino currents. If we consider the most
general case, the tree-level amplitude for the process can be
written in the form
\begin{eqnarray}
\label{amplitudeneutralcurrent1}
M=g\left({J_N}^\alpha{J'_N}_\alpha+{J_C}^\alpha\;{J_C}_\alpha\right)\mid_{m_\nu=0}
\end{eqnarray}
where ${J_N}^\alpha$ and ${J'_N}^\alpha$ are the neutral currents
for the neutrino and charged fermion respectively. ${J_C}^\alpha$ is
the charged neutrino current defined, similar to
(\ref{chargedcurrent-zeromass1}) and
(\ref{chargedcurrent-zeromass2}).  For completeness, let us write
the explicit form of the neutral neutrino current:
\begin{eqnarray}
\label{neutralcurrent-zeromass} {J_N}^\alpha&&=\left[\bar u_{\nu_f}
\gamma^\alpha \hat L u_{\nu_i}\right]
\end{eqnarray}
Here, $u_{\nu_i}$ ($u_{\nu_f}$) represents the spinor for initial
(final) neutrino field. Similar to the single neutrino absorption,
we do not perform an average over initial neutrino spins and omit
the factor of 1/2 coming from spin average of initial neutrinos.

\subsection{Massive case without mixing}

Now assume that neutrino masses are not strictly zero although they
are extremely small. We also assume that flavor and mass eigenstates
of the neutrino coincide, i.e., we ignore the mixing. Then, the
polarized cross section for a process where the neutrino has a spin
orientation defined by the proper angle $\theta$, can be obtained by
inserting the spinors (\ref{spinup}) or (\ref{spindown}) into the
relevant squared amplitudes and performing the phase space
integration. We additionally assume that the energy scale of the
process is much greater than the mass of the neutrino, $E>>m_\nu$.
Then, it is a very good approximation to use the expressions
obtained in the $m_\nu\to0$ limit. Therefore, during calculations,
the zero-mass limit of the spinors (\ref{limitspinup}) and
(\ref{limitspindown}) can be used instead of (\ref{spinup}) and
(\ref{spindown}).

If we insert the spinors $u^{(\uparrow)}(p)$ and
$u^{(\downarrow)}(p)$ describing a general spin orientation (spin
orientation defined by the proper angle $\theta$) into charged
neutrino current and take the $m_\nu\to0$ limit, we obtain
\begin{eqnarray}
\label{chargedcurrent-spinup} \lim_{m_\nu\to
0}{J^{(\uparrow)}_C}^\alpha&&=\sin{\left(\frac{\theta}{2}\right)}\left({J_C}^\alpha\mid_{m_\nu=0}\right)\\
\label{chargedcurrent-spindown} \lim_{m_\nu\to
0}{J^{(\downarrow)}_C}^\alpha&&=\cos{\left(\frac{\theta}{2}\right)}\left({J_C}^\alpha\mid_{m_\nu=0}\right)
\end{eqnarray}
where ${J_C}^\alpha\mid_{m_\nu=0}$ is the charged current for
massless neutrinos defined in (\ref{chargedcurrent-zeromass1}) or
(\ref{chargedcurrent-zeromass2}). While calculating the spin-up and
spin-down neutrino currents in the above equations, we make use of
the following identities: $\hat L\left\{\lim_{m_\nu\to
0}u^{(+)}(p)\right\}=\hat L u^{(R)}(p)=0$ and $\hat
L\left\{\lim_{m_\nu\to 0}u^{(-)}(p)\right\}=\hat L
u^{(L)}(p)=u^{(L)}(p)$. We also use the continuity of the helicity
states at $m_\nu=0$: $\lim_{m_\nu\to
0}u^{(+,-)}(p)=u^{(+,-)}(p)\mid_{m_\nu=0}=u^{(R,L)}(p)$. The squared
amplitude for single neutrino production or absorption processes
$ab\to \nu c$ or $\nu a'\to b'c'$ discussed in the previous
subsection is then found to be
\begin{eqnarray}
\label{squaredamplitude1} \lim_{m_\nu\to
0}|M^{(\lambda)}|^2=\frac{\left(1-\lambda\cos\theta\right)}{2}\left(|M|^2\mid_{m_\nu=0}\right)
\end{eqnarray}
where $|M|^2\mid_{m_\nu=0}$ is the squared amplitude for massless
neutrinos and $\lambda=+1$ corresponds to spin-up ($\uparrow$) and
$\lambda=-1$ corresponds to spin-down ($\downarrow$) polarization.
We observe from (\ref{squaredamplitude1}) that the squared amplitude
and consequently the cross section has a discontinuity at $m_\nu=0$.
For instance, if we choose $\theta=\pi/2$ (transverse polarization)
$m_\nu\to0$ limit of the cross section gives half of the cross
section for massless neutrinos: $\lim_{m_\nu\to
0}\sigma^{(\lambda)}\mid_{\theta=\pi/2}=\frac
{1}{2}\left(\sigma\mid_{m_\nu=0}\right)$. The cross section for
transverse polarization remains finite in the $m_\nu\to0$ limit but
it vanishes instantly at the point $m_\nu=0$. Let us examine the
zero-mass behavior of the cross section when the neutrinos are
described by helicity states. The negative (positive) helicity
corresponds to the choice $\lambda=-1$ and $\theta=0$ ($\lambda=+1$
and $\theta=0$). We see from (\ref{squaredamplitude1}) that the
cross section for positive helicity goes to zero and the cross
section for negative helicity goes to $\sigma\mid_{m_\nu=0}$ as
$m_\nu\to0$. Hence, if we restrict ourselves to special spin
orientations namely helicity states, we do not encounter any
discontinuity at $m_\nu=0$. However, the zero-mass continuity
observed from helicity states is misleading and does not hold true
in general as has been clearly shown above.

The longitudinal polarization of the neutrino is usually defined as
follows
\begin{eqnarray}
\label{longitudinalpolarization}
P_{\text{long}}=\frac{\sigma^{(+)}-\sigma^{(-)}}{\sigma^{(+)}+\sigma^{(-)}}
\end{eqnarray}
where $\sigma^{(+)}$ and $\sigma^{(-)}$ are the cross sections for
positive and negative helicity neutrinos. $P_{\text{long}}$ was
calculated for various SM processes in the literature (for example,
see Ref. \cite{Barenboim:1996cu}). It was shown that
$P_{\text{long}}$ goes to $-1$ as the neutrino mass approaches zero.
Indeed as we have discussed in the previous paragraph, according to
the squared amplitude (\ref{squaredamplitude1}),
$\lim_{m_\nu\to0}\sigma^{(+)}=0\Rightarrow
\lim_{m_\nu\to0}P_{\text{long}}=-1$. However, it is not correct to
conclude from this result that the neutrinos become completely
longitudinally polarized and described by 100\% negative helicity
states in the $m_\nu\to0$ limit. This is evident since the helicity
basis is not the only basis that spans the Hilbert space of the spin
states. A transversely polarized state is given by the superposition
of positive and negative helicity states and vanishing of the
positive helicity does not require the transverse polarization to be
zero. As we have discussed, although the cross section for positive
helicity goes to zero as $m_\nu\to0$, the cross section for
transverse polarization does not go to zero, instead it approaches
half of the cross section for massless neutrinos in that limit.
Therefore, the quantity $P_{\text{long}}$ defined in
(\ref{longitudinalpolarization}) is not the genuine measure of the
longitudinal polarization. It measures only the asymmetry between
positive and negative helicity states. If we define the quantity
which we call the degree of transverse polarization by
\begin{eqnarray}
\label{genuinelongitudinalpolarization}
P_{\text{trans}}=\frac{\sigma^{(T)}}{\sigma^{(+)}+\sigma^{(-)}}
\end{eqnarray}
we deduce that $\lim_{m_\nu\to0}P_{\text{trans}}=1/2$. Here,
$\sigma^{(T)}$ represents the cross section for either spin-up
($\lambda=+1$ and $\theta=\pi/2$) or spin-down ($\lambda=-1$ and
$\theta=\pi/2$) state of the transverse polarization.

The polarized cross section for neutrino scattering process $\nu
a\to \nu a$ can be calculated in a similar manner. If we insert the
spinors for a general spin orientation into neutral neutrino current
and take the $m_\nu\to0$ limit, we obtain
\begin{eqnarray}
\label{neutralcurrent-spin} \lim_{m_\nu\to
0}{J^{(\lambda_i,\lambda_f)}_N}^\alpha&&=\left[\frac{\left(1-\lambda_i\cos\theta_i\right)}{2}\right]^{1/2}
\left[\frac{\left(1-\lambda_f\cos\theta_f\right)}{2}\right]^{1/2}\left({J_N}^\alpha\mid_{m_\nu=0}\right)
\end{eqnarray}
where $\lambda_i$ and $\theta_i$ ($\lambda_f$ and $\theta_f$) belong
to the initial state (final state) neutrino and
${J_N}^\alpha\mid_{m_\nu=0}$ is the neutral current for massless
neutrinos defined in (\ref{neutralcurrent-zeromass}). The squared
amplitude for neutrino scattering process $\nu a\to \nu a$ is then
found to be
\begin{eqnarray}
\label{squaredamplitude2} \lim_{m_\nu\to
0}|M^{(\lambda_i,\lambda_f)}|^2=\left(\frac{1-\lambda_i\cos\theta_i}{2}\right)\left(\frac{1-\lambda_f\cos\theta_f}{2}\right)\left(|M|^2\mid_{m_\nu=0}\right)
\end{eqnarray}
where $|M|^2\mid_{m_\nu=0}$ is the squared amplitude for massless
neutrinos. The cross section of the neutrino-electron scattering for
polarized initial state neutrinos with general spin orientation and
unpolarized final state neutrinos, was calculated in
Ref.\cite{Kayser}. To obtain the cross section for unpolarized final
state neutrinos we should sum the squared amplitude over
$\lambda_f$, which gives:
\begin{eqnarray}
\label{amplitudeKayser} \lim_{m_\nu\to
0}|M^{(\lambda_i)}|^2=\sum_{\lambda_f=+1,-1}\left\{\lim_{m_\nu\to
0}|M^{(\lambda_i,\lambda_f)}|^2\right\}=\left(\frac{1-\lambda_i\cos\theta_i}{2}\right)\left(|M|^2\mid_{m_\nu=0}\right).
\end{eqnarray}
This squared amplitude coincides with the result of
Ref.\cite{Kayser} with only one difference that $m_\nu\to 0$ limit
in the left-hand side of (\ref{amplitudeKayser}) appears in our
calculations but it is absent in Ref.\cite{Kayser}. Instead,
spin-dependent squared amplitude was evaluated at $m_\nu=0$, i.e.,
according to \cite{Kayser}:
$|M^{(\lambda_i)}|^2\mid_{m_\nu=0}=\left(\frac{1-\lambda_i\cos\theta_i}{2}\right)\left(|M|^2\mid_{m_\nu=0}\right)$.
It seems the authors assumed that the spinors for a general spin
orientation have a continues behavior in the massless limit, i.e.,
they assumed $\lim_{m\to0}u^{(s)}(p)= u^{(s)}(p)\mid_{m=0}$. We also
would like to draw reader's attention to the following point. We see
from equations (\ref{squaredamplitude1}), (\ref{squaredamplitude2})
and (\ref{amplitudeKayser}) that the perpendicular component
(relative to momentum direction) of the spin three-vector $\vec n$
does not appear in the squared amplitudes. Recall that we choose
$\vec{n}=\sin\theta\; \hat x+ \cos\theta\; \hat z$ and $\vec p=p\hat
z$. Therefore $\vec n=(n_\perp,0,n_\shortparallel)$ where
$n_\perp=\sin\theta$ and $n_\shortparallel=\cos\theta$. Then
equation (\ref{amplitudeKayser}) can be written as $\lim_{m_\nu\to
0}|M^{(\lambda_i)}|^2=\left(\frac{1-\lambda_in_\shortparallel}{2}\right)\left(|M|^2\mid_{m_\nu=0}\right)$.
However, the disappearance of $n_\perp$ in the squared amplitude
does not imply that the squared amplitude is independent from
$n_\perp$. This is obvious because we have a condition between
$n_\perp$ and $n_\shortparallel$ obtained from the normalization of
the spin four-vector $s^\mu s_\mu=-1\Rightarrow\vec n\cdot \vec
n=n_\perp^2+n_\shortparallel^2=1$. Therefore we have one independent
parameter representing the orientation of the spin. One may decide
to choose $n_\perp$ or $n_\shortparallel$ as an independent
parameter or for instance, the proper angle $\theta$ as we did in
this paper. Regardless of which parameter we choose, the cross
section for transverse polarization evaluated in the $m_\nu\to0$
limit gives half of the cross section for massless neutrinos:
$n_\perp=1\Rightarrow n_\shortparallel=0\Rightarrow\lim_{m_\nu\to
0}\sigma=\frac{1}{2}\left(\sigma\mid_{m_\nu=0}\right)$. Thus, the
production, absorption and scattering probability of the neutrinos
with transverse polarization cannot be neglected.

\begin{figure}
\includegraphics[scale=1]{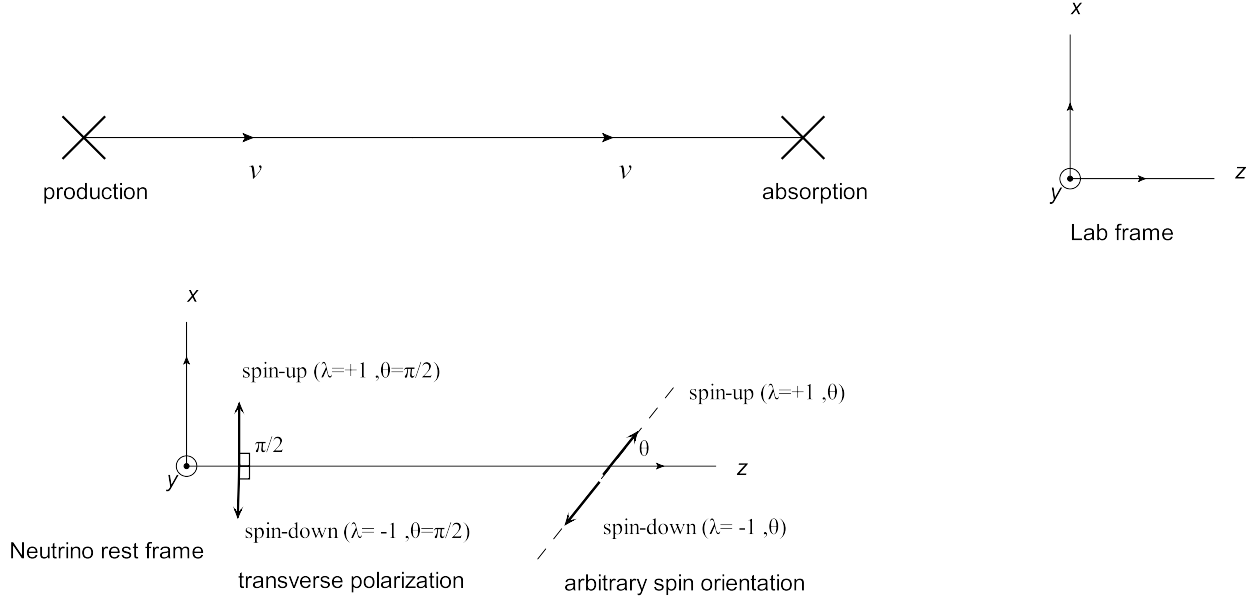}
\caption{ A schematic diagram which shows transverse and arbitrary
spin orientations in the neutrino rest frame, and neutrino
production and absorption observed in the laboratory
frame.\label{fig1}}
\end{figure}

The zero-mass discontinuity that we have discussed has important
implications on neutrino physics. It makes a significant distinction
between the cases in which neutrinos are exactly massless and
neutrinos have non-zero but very tiny masses. In the former case,
all SM neutrinos are described by completely longitudinally
polarized negative helicity states. Therefore, the factor $1/2$ due
to spin average of initial state neutrinos is omitted for processes
where neutrinos take part in the initial state. However, in the
later case it is not possible anymore to assume that neutrinos are
completely longitudinally polarized. This is obvious because, the
production cross section and hence the production probability of the
neutrinos with transverse spin orientation through SM processes
cannot be neglected. Therefore some part of the neutrinos in the SM
is transversely polarized. Consequently, the spin average of initial
state neutrinos in a process cannot be neglected and the cross
section is reduced due to this spin average. Our reasoning can be
presented in detail as follows: Consider a process in which the
neutrinos take part in the initial state. For example it might be
the neutrino absorption or scattering process. In order to calculate
the unpolarized total cross section we have to average over initial
and sum over final state spins. Some of the initial state neutrinos
are transversely polarized. Therefore for these neutrinos, spin
average is performed over spin-up and spin-down states of the
transverse polarization (FIG.\ref{fig1}). Then in the $m_\nu\to0$
limit, the unpolarized cross section gives
\begin{eqnarray}
\label{unpolcrosssection1} \lim_{m_\nu\to
0}\sigma^{(\text{unpol})}=\frac{1}{2}\sum_{\lambda_i=+1,-1}\lim_{m_\nu\to
0}\sigma^{(\lambda_i)}=\frac{1}{2}\left(\sigma\mid_{m_\nu=0}\right)
\end{eqnarray}
where we use $\lim_{m_\nu\to
0}\sigma^{(\lambda_i=+1)}=\lim_{m_\nu\to
0}\sigma^{(\lambda_i=-1)}=\frac{1}{2}\left(\sigma\mid_{m_\nu=0}\right)$
for $\theta=\pi/2$ (transverse polarization). We see from equation
(\ref{unpolcrosssection1}) that the unpolarized cross section is
reduced by a factor of $1/2$ compared to the cross section for
massless neutrinos. Hence, for transversely polarized initial state
neutrinos we obtain an average factor of $1/2$. However, not all
initial state neutrinos are transversely polarized. Some others are
longitudinally polarized. Since the cross section for neutrinos with
positive helicity is zero in the zero-mass limit, longitudinally
polarized initial neutrino states consist of $100\%$ negative
helicity states. In this case we do not perform an average over
initial neutrino spins and the unpolarized cross section is equal
the cross section for massless neutrinos:
\begin{eqnarray}
\label{unpolcrosssection2} \lim_{m_\nu\to
0}\sigma^{(\text{unpol})}=\lim_{m_\nu\to
0}\sigma^{(-)}=\left(\sigma\mid_{m_\nu=0}\right).
\end{eqnarray}
We have deduced from the above analysis that if we consider
transversely polarized part of the initial neutrinos, then $50\%$ of
them are spin-up and other $50\%$ are spin-down. We should then
perform an average over initial spins which gives a factor of $1/2$.
On the other hand, if we consider longitudinally polarized part of
the initial neutrinos, then $100\%$ of them are negative helicity
and none of them are positive helicity. Then we do not perform an
average and instead of $1/2$ we get a factor of $1$. Hence, an
important question arises:  By which factor is the cross section
reduced? In order to give an answer to this question, let us
consider the following gedankenexperiment. Assume that neutrinos are
detected in a particle detector via the absorption process $\nu
a'\to b'c'$. Without loss of generality, also assume that all the
detected neutrinos, are produced via the production processes $ab\to
\nu c$. FIG.\ref{fig1} represents a schematic diagram for this
gedankenexperiment.
We will use the subscript "1" to denote the production process
$ab\to \nu c$ and subscript "2" to denote the absorption process
$\nu a'\to b'c'$. If the neutrinos are produced in a particle
accelerator then the total number of produced neutrinos is given by
$N_1=\sigma_1^{(\text{unpol})}\;L_1$, where
$\sigma_1^{(\text{unpol})}$ is the unpolarized total cross section
and $L_1$ is the integrated luminosity. If we assume that \emph{all}
of the produced neutrinos have a fix spin orientation defined by the
proper angle $\theta$, then the number of produced neutrinos with
this spin orientation is given by
\begin{eqnarray}
\label{neventproduced}
N_1^{(\lambda)}(\theta)=\sigma_1^{(\lambda)}(\theta)\;L_1=\frac{\left(1-\lambda\cos\theta\right)}{2}\left(\sigma_1\mid_{m_\nu=0}\right)\;L_1
\end{eqnarray}
where we take $m_\nu\to0$ limit and make use of equation
(\ref{squaredamplitude1}). We observe from (\ref{neventproduced})
that the total number of produced neutrinos is independent from the
proper angle $\theta$:
\begin{eqnarray}
\label{neventproduced_total}
N_1=N_1^{(\lambda=+1)}(\theta)+N_1^{(\lambda=-1)}(\theta)=\left(\sigma_1\mid_{m_\nu=0}\right)\;L_1.
\end{eqnarray}
Now we consider a massive detector which is composed of a huge
number of atoms. The produced neutrinos can interact with the
electrons and nucleons (or quarks) of the detector through the
process $\nu a'\to b'c'$ and the detection occurs. For simplicity we
assume that all of the produced neutrinos are passing through the
detector. Then, the number of detected neutrinos with spin
orientation defined by the proper angle $\theta$ can be written as
\begin{eqnarray}
\label{neventdetected}
N_2^{(\lambda)}(\theta)=N_1^{(\lambda)}(\theta)P^{(\lambda)}(\theta)
=\sigma_1^{(\lambda)}(\theta)\sigma_2^{(\lambda)}(\theta)\;L_1L_2.
\end{eqnarray}
Here $P^{(\lambda)}(\theta)=\sigma_2^{(\lambda)}(\theta)L_2$ is the
detection probability of a single polarized neutrino and $L_2$ is a
constant that depends on the parameters of the detector. For
instance, $L_2$ depends on the number of electrons or nucleons per
unit volume, the fiducial mass of the detector, etc. Since the
details of the detector are irrelevant to our analysis, we do not
consider its explicit form as a function of detector parameters and
assume that it is just a constant. According to equation
(\ref{squaredamplitude1}) the zero-mass limit of the production and
absorption cross sections can be written as
$\lim_{m_\nu\to0}\sigma_{1,2}^{(\lambda)}(\theta)=\frac{\left(1-\lambda\cos\theta\right)}{2}\left(\sigma_{1,2}\mid_{m_\nu=0}\right)$.
The total number of detected neutrinos is then
\begin{eqnarray}
\label{neventdetected_total}
N_2^{(\lambda=+1)}(\theta)+N_2^{(\lambda=-1)}(\theta)=\left[\sin^4\left(\frac{\theta}{2}\right)
+\cos^4\left(\frac{\theta}{2}\right)\right]N_2.
\end{eqnarray}
where $N_2=\left(\sigma_1\mid_{m_\nu=0}\right)
\left(\sigma_2\mid_{m_\nu=0}\right) L_1L_2$ is the total number of
detected neutrinos in case all produced neutrinos are massless. In
the left hand side of (\ref{neventdetected_total}), the limit
${m_\nu\to0}$ is implemented but not shown.

In the above analysis we assume that all the neutrinos are produced
having the same spin orientation with respect to the direction of
momentum, i.e., with a same proper angle $\theta$. Specifically if
we assume that all produced neutrinos are transversely polarized
($\theta=\pi/2$) then total number of detected neutrinos is $N_2/2$.
On the other hand, if all produced neutrinos are longitudinally
polarized ($\theta=0$) then total number of detected neutrinos is
$N_2$. However in a real situation, the produced beam is comprised
from neutrinos with different spin orientations. Hence, we should
consider every possible spin orientations and an average over
different spin orientations has to be performed. It is easy to show
that the average of the trigonometric expressions in the square
parentheses yields $\langle
\left[\sin^4\left(\frac{\theta}{2}\right)
+\cos^4\left(\frac{\theta}{2}\right)\right]\rangle=2/3$. Here we
should note that a statistical weight of $\sin\theta$ is used during
the average. Therefore different from other standard model fermions,
spin average of initial state neutrino in a SM process yields a
factor of $2/3$ instead of $1/2$. Hence the total cross section is
reduced by this factor compared to the case in which neutrinos are
described by $100\%$ negative helicity states. Here we should
emphasize that the standard model fermions other than neutrinos
carry electric and/or color charge and they interact dominantly
through vector type coupling. Since the vector coupling does not
provide a preferred spin orientation, all different orientations of
their spin three-vector $\vec n$ are equally probable unless they
are intentionally produced polarized. Therefore, for initial state
electrons, quarks, etc. the average over proper angle $\theta$ is
omitted. On the other hand, the average over spin-up and spin-down
states is performed and yields a factor of $1/2$.

Let's summarize what we have done so far: We have deduced that due
to zero-mass discontinuity in the cross section the cases in which
neutrinos are exactly massless and neutrinos have non-zero but very
tiny masses, have completely different implications. Therefore,
contrary to the previously accepted opinion in the literature, it is
not a good approximation to neglect neutrino masses during cross
section calculations even though neutrino masses are very small and
the energy scale of the processes are much greater than the neutrino
mass. \emph{We have deduced a surprising result that the total cross
section of the process where a neutrino takes part in the initial
state is reduced by a factor of 2/3 due to spin average.} As far as
we know, this fact has been overlooked in the literature. In the
previous studies on this subject, the spin average of initial state
neutrinos was omitted for processes where neutrinos take part in the
initial state.

The total neutrino (anti-neutrino) cross sections have been measured
in plenty number of experiments since the famous experiments of
Cowan and Reines \cite{Reinescowan1,Reinescowan2}. In all these
experiments the measured cross sections seem not to be reduced by
the factor $2/3$. They confirm the fact that neutrino states are
almost $100\%$ negative helicity. Possibly because of the
experimental verification of the neutrino helicity, theoretical
predictions have not been examined in much detail by previous
studies. However, as we have deduced, a straightforward calculation
taking into account the existing zero-mass discontinuity of the free
Dirac spinors yields a discrepancy between quantum field theory
predictions and the experimental results. One possible solution to
this problem might be provided by adding a new simple hypothesis to
established axioms of quantum field theory \cite{Sahin:2015ofl}. The
scope of this paper is limited; we do not aim to discuss possible
solutions to the discrepancy. Our purpose is just to reveal the
surprising consequences of the zero-mass discontinuity of the Dirac
spinors on neutrino cross sections.

In closing to this subsection, we would like to draw reader's
attention to another surprising consequence of the zero-mass
discontinuity of the Dirac spinors. Throughout this paper, all
calculations have been carried out considering only Dirac neutrinos.
It is assumed that Dirac and Majorana neutrino cross sections
coincide in the $m_\nu\to0$ limit \cite{Kayser,Barranco:2014cda}.
This fact is based on the assumption that both Dirac and Majorana
spinors become completely left-handed chiral in the $m_\nu\to0$
limit. However, as we have discussed in section II, a free Dirac
spinor with arbitrary spin orientation does not necessarily result
in a chirality eigenstate in the zero-mass limit. Therefore,
contrary to expectations, Dirac and Majorana cross sections can lead
to different results even though the limit $m_\nu\to0$ is performed.

\subsection{Massive case with mixing}

We have so far ignore the mixing between different mass eigenstates
of the neutrino. However, in a realistic situation the neutrinos
interact through weak interaction in flavor eigenstates which are
given by a superposition of the mass eigenstates. The mixing
equation is given by $\nu_{\ell L}=\sum_{i=1}^3U_{\ell i}\;\nu_{i
L}$  where  $U_{\ell i}$ is the Pontecorvo-Maki-Nakagawa-Sakata
(PMNS) matrix element \cite{Bilenky,Kayserbook}. Here we use the
subscript $\ell$ to denote the flavor and subscript $i$ to denote
the mass eigenstates. Therefore, the scattering process for
$\nu_\ell$ consist of separate processes for mass eigenstates
$\nu_i$, $i=1,2,3$. The cross section calculations are then
performed for neutrino mass eigenstates and the contributions coming
from different mass eigenstates are added. According to the minimal
extension of the SM with massive neutrinos, the scattering amplitude
for $\nu_i$ is almost same with the amplitude for a neutrino without
mixing. The only difference is that the charged neutrino current
picks up an extra factor $U_{\ell i}$. It is obvious that the
surprising result that we encounter in the previous subsection is
also true when we consider the processes $ab\to\nu_ic$, $\nu_ia'\to
b'c'$ and $\nu_ia\to \nu_ia$ where the neutrinos are taken to be in
the mass eigenstate. The neutrino mixing does not solve the problem,
on the contrary, the problem becomes worse than it was before. Since
various different spin orientations of the neutrino contribute to
the cross section, we can conceive the flavor eigenstate as a
superposition of the mass eigenstates where each mass eigenstate may
have an arbitrary spin orientation. Then, the spin state of the
flavor eigenstate becomes ambiguous. One may assume the flavor
neutrino has a mixed spin state, in the sense that, each of its
constituent mass eigenstates has a different spin orientation. Let
us consider the single neutrino production or absorbtion processes
discussed in the previous subsections. If we sum the squared
amplitudes that belong to individual mass eigenstates we expect to
obtain the squared amplitude for the flavor eigenstate:
$|M_\ell|^2=\sum_i|M_i|^2$. According to equation
(\ref{squaredamplitude1}) the sum over mass eigenstates gives:
\begin{eqnarray}
\lim_{m_\nu\to0}\sum_i|M^{(\lambda_i)}_i|^2=\sum_i\left[|U_{\ell
i}|^2\frac{\left(1-\lambda_i\cos\theta_i\right)}{2}\right]\left(|M_\ell|^2\mid_{m_\nu=0}\right)
\end{eqnarray}
where $\left(|M_\ell|^2\mid_{m_\nu=0}\right)$ is the squared
amplitude for the flavor neutrino evaluated at $m_\nu=0$. In case
all spin orientations of the mass eigenstates are equal
($\lambda_1=\lambda_2=\lambda_3; \theta_1=\theta_2=\theta_3$), we
obtain the expected result:
\begin{eqnarray}
\lim_{m_\nu\to0}\sum_i|M^{(\lambda_i)}_i|^2=\frac{\left(1-\lambda\cos\theta\right)}{2}\left(|M_\ell|^2\mid_{m_\nu=0}\right)=\lim_{m_\nu\to0}|M^{(\lambda)}_\ell|^2
\end{eqnarray}
where we use the unitarity of the PMNS matrix. However, we do not
have any reasonable explanation for the choice
$\lambda_1=\lambda_2=\lambda_3; \theta_1=\theta_2=\theta_3$. In
general, spin orientations of different mass eigenstates can be
different.

\section{Conclusions}

The helicity states have a continuous behavior in the massless
limit. When we take $m\to0$ limit, a helicity state converge to one
of the chirality eigenstate and becomes completely left-handed or
right-handed chiral. The zero-mass behavior observed from helicity
states can make one think that massless limit is always smooth.
However, this behavior is specific to helicity states and is not
valid in general. Massless limit has some subtleties in the case of
spinors with general spin orientations. The angle which defines the
spin orientation of a fermion is an invariant quantity by
definition. Hence, the spin orientation of a fermion does not
necessarily becomes parallel or anti-parallel to the momentum
direction and does not necessarily result in a chirality eigenstate
in the zero-mass limit. This behavior makes free Dirac solutions
discontinues at $m=0$. We explore the consequences of this zero-mass
discontinuity of the Dirac spinors and show that it has surprising
consequences for neutrino cross sections.

The most challenging consequence of the zero-mass discontinuity is
that it yields a discrepancy between theoretical predictions and the
experimental results. We call this discrepancy the neutrino helicity
problem. The theoretical predictions of the cross section for
massive neutrinos with general spin orientation have been discussed
for decades. In this respect, many of the calculations presented in
this paper is not totally novel; the new idea of the paper, lies in
the reinterpretation of the dependence of the cross section on the
spin three-vector $\vec n$. Although the resultant discrepancy is
very disturbing, we decide to present our results since we think
that they are concrete predictions of the theory. The neutrino
helicity problem points out that something is wrong in the
assumptions used in the theory. The polarized cross section
calculation technique for a general spin orientation is a
conventional method which is used successfully for other fermions.
Indeed, top quark spin polarization has been measured for various
spin orientations and it was found to be consistent with the
theoretical predictions \cite{Abazov:2016tba,Abazov:2015fna}.
Therefore, the problem should be associated with the neutrino
nature.

\begin{acknowledgments}
The author thanks Prof. A. U. Y{\i}lmazer for helpful criticism of
the manuscript and valuable suggestions.
\end{acknowledgments}

\end{document}